\documentclass[fleqn,usenatbib]{mnras}

\usepackage[T1]{fontenc}
\usepackage{multirow}
\usepackage{amsmath, bm}
\usepackage{float}
\usepackage{soul}
\usepackage{booktabs}
\DeclareRobustCommand{\VAN}[3]{#2}
\let\VANthebibliography\thebibliography
\def\thebibliography{\DeclareRobustCommand{\VAN}[3]{##3}\VANthebibliography}

\usepackage{graphicx}	
\usepackage{amsmath}	
\usepackage{amssymb}	

\title[Galaxy Morphology Classification Using MSCCN]{Galaxy Morphology Classification Using Multi-Scale Convolution Capsule Network}

\author[G. P. LI et al.]{
Guangping Li,$^{1}$\thanks{E-mail: 15136131592@163.com}
Tingting Xu,$^{1}$
Liping Li,$^{1}$
Xianjun Gao,$^{1}$
Zhijing Liu,$^{1}$
Jie Cao,$^{1}$
\newauthor
Mingcun Yang,$^{1}$
Weihong Zhou,$^{1,2}$\thanks{E-mail:ynzwh@163.com}
\\
$^{1}$School of Mathematics and Computer Science, Yunnan Minzu University, Kunming 650504,China\\
$^{2}$Key Laboratory for the Structure and Evolution of Celestial Objects, Chinese Academy China of Sciences, Kunming 650011, China }

\date{Accepted XXX. Received YYY; in original form ZZZ}

\pubyear{2023}

\begin{document}
\label{firstpage}
\pagerange{\pageref{firstpage}--\pageref{lastpage}}
\maketitle

\begin{abstract}
The classification of galaxy morphology is a hot issue in astronomical research. Although significant progress has been made in the last decade in classifying galaxy morphology using deep learning technology, there are still some deficiencies in spatial feature representation and classification accuracy.
In this study, we present a multi-scale convolutional capsule network (MSCCN) model for the classification of galaxy morphology. 
First, this model improves the convolutional layers through using a multi-branch structure to extract multi-scale hidden features of galaxy images. In order to further explore the hidden information in the features, the multi-scale features are encapsulated and fed into the capsule layer. Second, we use a sigmoid function to replace the softmax function in dynamic routing, which can enhance the robustness of MSCCN.
Finally, the classification model achieving 97\% accuracy, 96\% precision, 98\% recall, and 97\% F1-score under macroscopic averaging.
In addition, a more comprehensive model evaluation were accomplished in this study. We visualized the morphological features for the part of sample set, which using the t-distributed stochastic neighbor embedding (t-SNE) algorithm. 
The results shows that the model has the better generalization ability and robustness, it can be effectively used in the galaxy morphological classification.
\end{abstract}




\section{Introduction}
The galaxy is a celestial system composed of star and interstellar medium with a spatial scale of thousands to hundreds of thousands of light-years, which includes environmental density, interact history, gas accretion, and dark matter halo \citep{8193490}. 
Most galaxies have obvious geometric features. The main structures of the galaxy are spheres and disks, and the rods and the spiral arms in disks \citep{2003ApJ...588..218A}. Through the evolution of galaxy structure with redshift, we can understand the formation and evolution of galaxy \citep{Wang2007Progressinastronomy}. In this study, the way to classify a galaxy accurately and to understand its morphological structure is of great importance for studying the physical properties of the galaxy.

According to different classification criteria, the galaxy has different classification systems. In the early stage, limited by observation equipment, acquisition technology and other factors, only some bright and near galaxy images could be obtained and directly classified by human eyes, which is called the visual classification system. 
The typical method is Hubble classification system \citep{1926ApJ....64..321H}. According to their morphology, which divides galaxies into four basic types: spiral galaxy ($\mathit{S}$), barred spiral galaxy ($\mathit{SB}$), elliptical galaxy ($ \mathit{E}$), irregular ($\mathit{Irr}$). 
Since then, the modeling classification systems and non-model classification systems have been proposed. The modeling classification system is mainly based on different galaxy morphology corresponding to different surface brightness profiles to divide the morphology of galaxy.
The most common methods are exponential 
fitting \citep{sersic1968atlas} and nuclear bulge-nuclear disk fitting \citep{1998AJ....116.2644O}. 
According to structural parameters such as clumpiness index \citep{2003ApJS..147....1C}, moment index \citep{2004AJ....128..163L} and concentration index \citep{bershady2000structural}, the non-model galaxy classification system mainly classifies galaxy morphology.

Since the new century, the Sloan Digital Sky Survey (SDSS) \citep{lupton2001sdss}, Large Synoptic Survey Telescope (LSST) \citep{2019ApJ...873..111I}, Cosmic Evolution Survey (COSMOS) \citep{scoville2007cosmic}, LAMOST \citep{zhao2012lamost}, Space Infrared Telescope Facility (SIRTF) \citep{1998SPIE.3356..478F}, and James Webb Space Telescope (JWST) \citep{gardner2006james} are gradually implemented. 
Traditional data processing methods are difficult to process the galaxy data efficiently and real-time. 
The traditional galaxy morphology classification method distinguishes different classes of galaxy by manually extracting galaxy image features, including surface brightness ($\mathit{I(r)}$), Gini coefficien ($\mathit{G}$), semi-diameter ($\mathit{SD}$), and concentration \citep{sorrentino2006galaxy,2003ApJ...588..218A}. These methods depend on the comprehensiveness of feature extraction, besides the process of the selection and the extraction of the features is time-consuming and laborious. 

In recent years, due to its advantages such as the manual design features are not required, strong learning ability and hidden features are fully mining, deep learning has been widely used in galaxy morphology classification. 
\citet{zhu2019galaxy} proposed an improved deep residual network model to classify galaxy morphological, this model improves the residual unit, reduces the network depth, and widens the network width.
\citet{Gupta2022} introduced a continuous depth version of the Residual Network for the Galaxy Zoo2 (GZ2) \citep{hart2016galaxy} galaxy data set, it use Adjoint and Adaptive Checkpoint Adjoint (ACA) of numerical techniques to trains the Neural ordinary differential equations (NODE). Compared with the residual network, this model reduced about 1/3 of the parameters and achieved better effect of classification. 
\citet{GMC2022Wang} built a galaxy classification network (GMC-net) to classify the photometric images of galaxies in SDSS DR16, Galaxy Zoo2 and EFIGI catalog. The model can automatically extract the features of galaxy images and automatically classify them according to their shapes, avoiding the difficulties of the extraction and selection of features and selection of classifier.
\citet{mittal2020data} designed a convolutional neural network (CNN) classifier for the irregular galaxy, and this classifier has a data-enhanced function,
and the model uses certain data augmentation techniques and diffident activation functions to classify galaxy and obtain a better results than its earlier contemporaries. 
\citet{fielding2022classification} uses a convolutional auto-encoder as a feature extractor, and the features were clustered by k-means, fuzzy c-means and agglomerative clustering to classify the galaxy, this approach could better classifying Galaxy Zoo. 
\citet{2022RAA....22e5002Z} aims at the problems of large training data and parameters in supervised deep learning model, proposed a few-shot learning model based on Siamese Networks to classify galaxies, the model is not only suitable for the taxonomy of galaxy morphology but also for identifying rare astrophysical objects. Although the sample data is limited, the model still achieves excellent classification results. \citet{2021arXiv211001024Y} first applied Vision Transformer to the classification task of galaxy morphology and achieved fairly classification results with traditional CNNs, and the method is specifically good at classifying smaller-sized and fainter galaxies. \citet{nishikawa2020semi} proposed a semi-supervised deep learning method, this model uses a Variational Auto-encoder (VAE) with Equivariant Transformer layers as classifier to classify galaxy morphology. The novel method using the fewer labels data achieved a higher accuracy compared to exiting approaches.

However, neural networks that have been used previously for galaxy morphology classification have some disadvantages, among which CNN are representative. When it is used for galaxy morphology classification, the pooling operation in CNN will lead to some critical feature information loss. In addition, the CNN does not have the Isotropic, and only retains the size of feature but ignores the important information of direction and spatial position \citep{nishikawa2020semi}. 
These factors affect the classification performance of galaxy. In 2017, to solve the problems existing in CNN model, \citet{sabour2017dynamic} proposed a new deep learning network, which is called capsule network (CapsNet). The core unit of CapsNet is capsule, which is composed of a group of neurons, and it can store and output feature information in the form of vectors. The length of the capsule vector represents the probability of the object's existence and the direction represents the object's attribute (location,rotation, size, color, etc.). CapsNet are able to output more comprehensive feature information than CNNs that use only a single neuron for target representation.

Although the capsule network has many advantages in image classification, there are still some room to be improved, such as the insufficient ability of feature extraction, poor performance in some classification tasks, what's more, a large number of parameters and computations hinder the promotion of CapsNet. To solve the above problems, we propose a Multi-scale Convolutional Capsule Network (MSCCN) for the galaxy morphology classification.

The study is organised as follows: we discuss the tradition capsule network, and introduce the network structure of multi-scale convolutional neural networks in section \ref{sec:Methods}. In section \ref{sec:data set}, we introduce the data set and experiment equipment used in this study. Besides, we pre-process the data and select the best hyper-parameters for our model. In section \ref{sec:results}, we show the classification results of the model and compare our results with other similar works. In section \ref{sec:Visualization}, we visualized the output of the DigitCaps layer of the model, and analysis the physical representation of the low-level evidence of galaxy data. Finally, we present the conclusion in section \ref{sec:conclusion} .

\section{Methods}
\label{sec:Methods}

\subsection{Capsule Network}
\label{sec:Capsule Network}

\begin{figure}
	\includegraphics[width=\columnwidth]{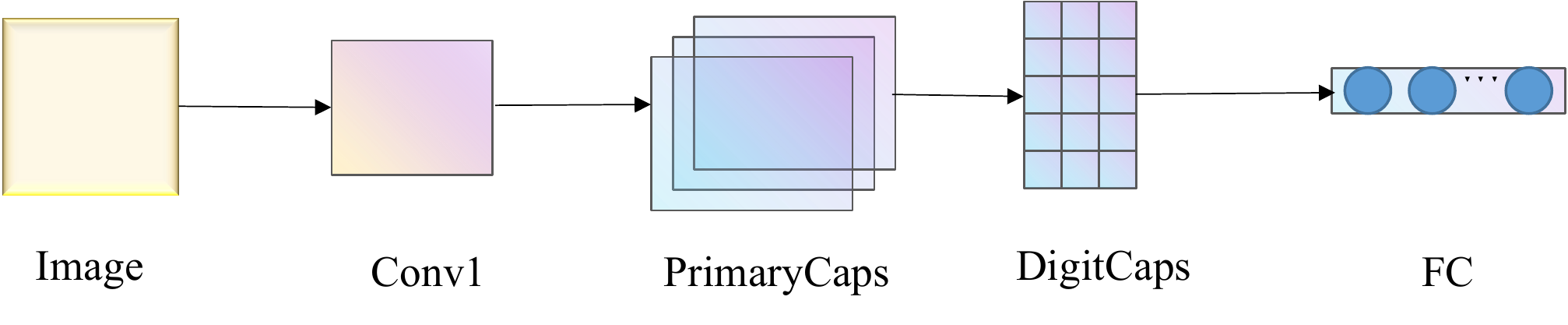}
	\caption{Structure of Traditional Capsule Networks}
	\label{fig:figure1}
\end{figure}

A complete CapsNet can be divided into encoder and decoder. The encoder contains the convolution layer, primary capsules (PrimaryCaps) layer and digital capsules (DigitCaps) layer. The decoder is composed of a full connection layer.
Furthermore, CapsNet can build spatial hierarchy through the dynamic routing process, and then extract spatial feature information of objects.
Dynamic routing is an algorithm based on protocol routing that predicts the capsule of the present layer through assigning a reasonable weight to each capsule of the last layer. Fig.\ref{fig:figure1} is the structure of traditional capsule network (CapsNet).

\subsubsection{Convolution Layer}

The convolution layer of CapsNet is used to extract the low-level features of the object. The convolution formula is defined as follows:

\begin{equation}
{y_{ij}} = \sum\limits_{u = 1}^m {\sum\limits_{v = 1}^n {{f_{uv}} \cdot {x_{i - u + 1,j - v + 1}}} }
\label{eq:quadratic}
\end{equation}
where the ${x_{ij}}, 1 \le i \le M, 1 \le j \le N$ is the image matrix of galaxy, ${f_{uv}},1 \le u \le m,1 \le v \le n$ is a filter. In the convolution layer, the input from the neuron $\mathit{i}$ to the layer $\mathit{l}$ is represented as: $a_i^l = f(\sum\nolimits_{j = 1}^m {w_j^{(l)} \cdot a_{i - j + m}^{(l - 1)} + {b^{(l)}}} )$. The $ \mathit{w_j^{(l)}} $ represents the convolution kernel, and $f$ represent the activation function. All neurons are same and the weights are shared in layer $\mathit{l}$.

\subsubsection{Primary Capsule Layer}
The primary capsule layer encapsulated primary features into vectors, and normalized galaxy data sets. We use the squashing function to normalized for the data, which has narrowed the input data interval, reduced the amount of network computation, and strengthened the ability of expression and learning of network. 
The expression of the primary capsule layer is

\begin{equation}
	{u^{l(i,j)}} = {f_s}\left( {\begin{array}{*{20}{c}}
			{{f_a}(\sum\limits_i {a_i^1)} }\\
			\vdots \\
			{{f_a}(\sum\limits_i {a_i^l} )}
	\end{array}} \right)
	\label{eq:quadratic}
\end{equation}
where the ${u^{l(i,j)}}$ represents the primary capsule, $f_{a}$ represents the operation of the primary capsule layer on feature data, ${f_a}(\sum\limits_i {a_i^l} )$ is the output of the convolution layer, and ${f_s}$ is the squash function, where it is nonlinear squash function
\begin{equation}
	{f_s}({f_a}(\sum\limits_i {a_i^l} )) = \frac{{{{\left\| {{f_a}(\sum\limits_i {a_i^l} )} \right\|}^2}}}{{1 + \left\| {{f_a}(\sum\limits_i {a_i^l} )} \right\|}} \cdot \frac{{{f_a}(\sum\limits_i {a_i^l} )}}{{\left\| {{f_a}(\sum\limits_i {a_i^l} )} \right\|}}.
	\label{eq:quadratic}
\end{equation}

\subsubsection{Digital Capsule Layer}
The information transmitted between the primary capsule layer and the digital capsule layer consists of two main processes: linear transformation and dynamic routing update.
In the linear transformation, an activation vector ${\hat u_{j|i}}$ can be obtained when low-level capsule $i$  transmits data to high-level capsule $j$ and is calculated by the equation:

\begin{equation}
    \hat u_{j|i}=w_{ij} \cdot u_i
    \label{eq4}
\end{equation}
Where, ${u_i}$ is output vectors of the low-level capsules, and ${w_{ij}}$ is an unique linear transformation matrix.

In the high-level capsule, the total input ${s_j}$ is obtained by calculating a weighted summation of the ${\hat u_{j|i}}$ and $c_{ij}$, and the process is expressed as
\begin{equation}
{s_j} = \sum\limits_{i = 1}^N {{c_{ij}} \cdot {\hat u_{j|i}}} 
	\label{eq:quadratic4}
\end{equation}
where, $c_{ij}$ is the coupling coefficients, which are determined by iterative dynamic routing. In the traditional CapsNet, $c_{ij}$ is calculated by the softmax function, and the calculation process is equation:

\begin{equation}
    {c_{ij}=\frac{exp(b_{ij})}{\sum\nolimits_k exp(b_{ik})}}
    \label{eq:quadratic6}
\end{equation}

\begin{figure}
	\includegraphics[width=\columnwidth]{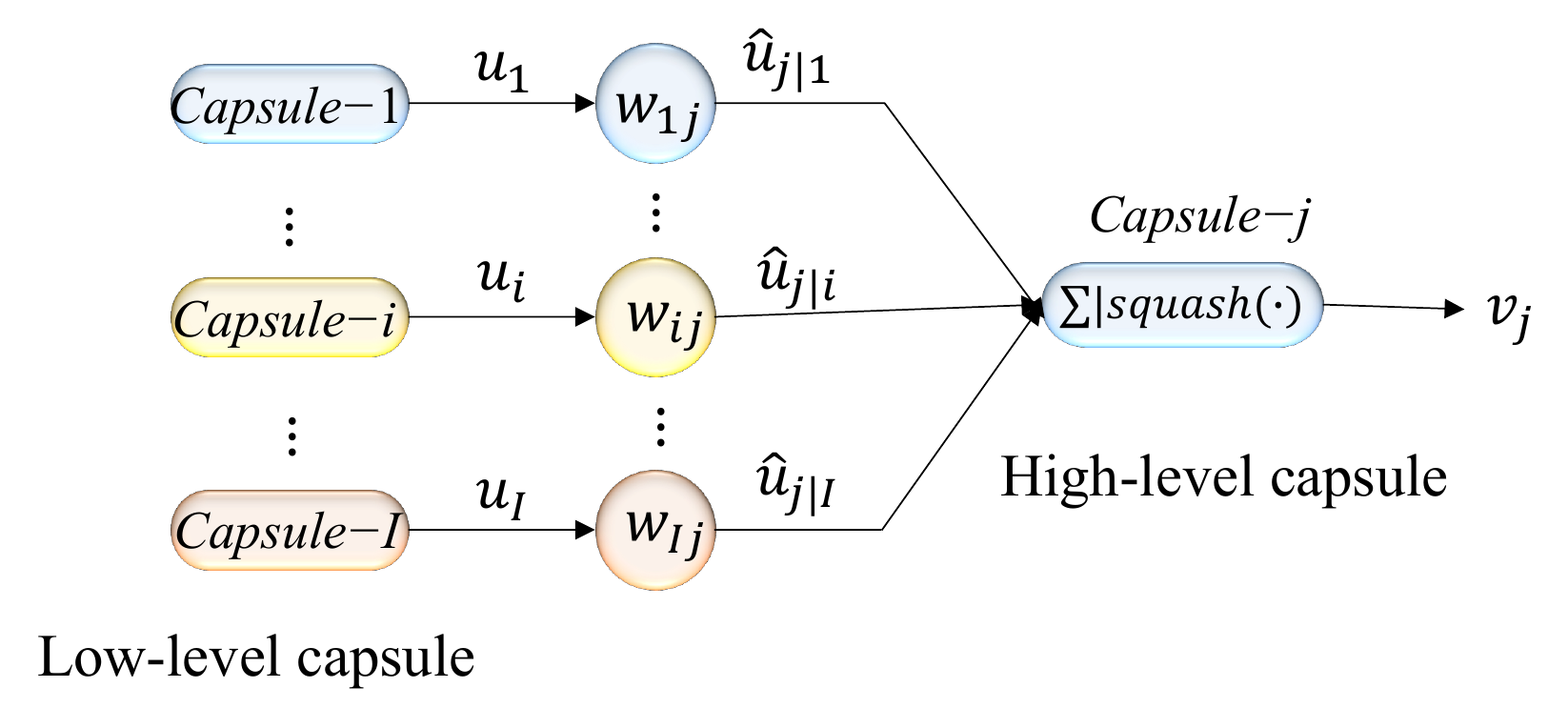}
	\caption{Update process of dynamic routing}
	\label{fig:example_figure}
\end{figure}

The parameter ${b_{ij}}={\hat u_{j|i}}\cdot{c_{ij}}$ is the priori probability of capsule $i$to capsule $j$. Through updating ${\hat u_{j|i}}$ and ${v_j}$ at first, and then ${b_{ij}}$, and $c_{ij}$ were updated too, this process is defined as a iteration of dynamic routing, Fig.\ref{fig:example_figure} present the updated process of dynamic routing. Lastly, a group of optimal parameters is obtained and the update of dynamic routing is completed.

Lastly, the output of the high-level capsule $j$ is noted as $v_{j}$, which can be calculated by the ${s_j}$. Meanwhile, the ${s_j}$ is squeezed through the squash function.

\begin{equation}
	{v_j} = \frac{{{{\left\| {{s_j}} \right\|}^2}}}{{1 + {{\left\| {{s_j}} \right\|}^2}}} \cdot \frac{{{s_j}}}{{\left\| {{s_j}} \right\|}}.
	\label{eq:quadratic}
\end{equation}

\subsubsection{Loss Function}
In this work, the total loss of the model is a combination of the margin loss and the reconstruction loss. In the digital capsule layer, margin losses are used to update the weights between digital capsules, which are calculated as follows:
\begin{equation}
	{L_k} = {T_k}\max (0,{m^ + } - {\left\| {{v_k}} \right\|^2}) + \lambda {(1 - {T_k})\max (0,\left\| {{v_k}} \right\| - {m^ - })^2}.
	\label{eq:quadratic}
\end{equation}
Where, the ${T_k}$ is a indicator function of classification. If the class of $k$ exists, then the ${T_k}$ is $1$; otherwise the ${T_k}$ is $0$. The ${m^+} = 0.9$ is false positive(FP), ${m^-} = 0.1$ is the false negative(FN). The $\lambda$ is defaulted to $0.5$.

The reconstruction loss represents the loss of the image reconstructed in decoder, it is the euclidean distance between the reconstructed vector and the input vector. In this work, we reduce this by a factor of 0.0005 to avoid the reconstruction loss dominating the margin loss.

There are three fully connected layers at the final part of the network. 
The feature space learned in the previous layers is mapped to the sample marker space in the three fully connected layers.
The galaxy morphological features learned in the previous layers are integrated, the galaxy image could be reconstructed.

\subsection{Multi-scale convolution capsule network}

\begin{figure*}
	\includegraphics[width=1.5\columnwidth]{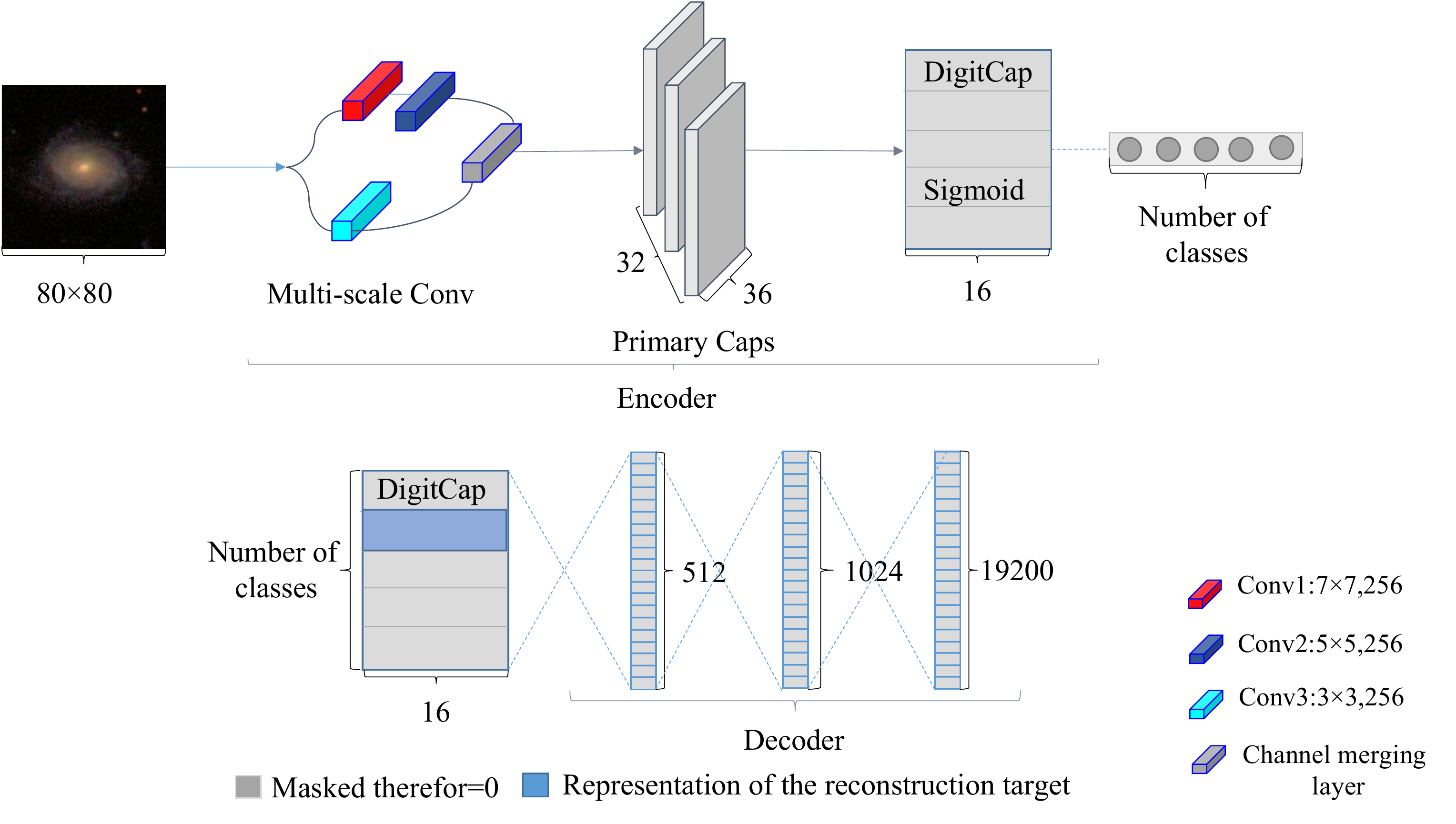}
	\caption{Structure of the MSCCN model. Inputs: $80\times80$ downsampled images of the galaxies. Sigmoid: A function is used to replace the softmax function in dynamic routing. Conv1: a convolutional layer with receptive filed of $7\times7$ and 256 filters. Conv2: a convolutional layer with receptive filed of $5\times5$ and 256 filters. Conv3: a convolutional layer with receptive filed of $3\times3$ and 256 filters. Conv4: Convergence of the features of the two channels. Primary Caps: The featurea extracted by convolutional layer reshaped to 32 primary caps with 8 dimension where each dimension is a feature map with the size of $32\times32$. DigitCaps: There are five capsules,each of them represent one class. Decoder: three full connected layers with 512, 1024, 19200 neurons, it is used to reconstructed the images of galaxy}.
	\label{fig:example_figure3}
\end{figure*}

In traditional CapsNet, the low-level feature extraction module has a simple structure that uses only a single convolution layer to extract the low-level features of the image. Its feature extraction activated is insufficient and the parameters are redundant. To further improve the performance of CapsNet, the multi-branch structure is used to improve the generation process of CapsNet capsules to increase the multi-scale feature extraction ability of model at first. Secondly, when the coupling coefficients are applied by the "routing softmax" function \citep{sabour2017dynamic}, the values of the coupling coefficients will be concentrated in a small interval, we use sigmoid function to replace the "routing softmax" function in dynamic routing, which can help the network get a more uniform distribution of routing coefficients and strengthen discernibility of the output vector for each class. Therefore, we constructed a multi-scale convolution capsule network model for the classification of the galaxy image. Fig.\ref{fig:example_figure3} is the model structure of the MSCCN network. The following contributions are presented in this work:

(1) Instead of convolution layer in CapsNet, we use a multi-scale parallel convolution layer to extract multi-scale hidden features of galaxy images.

(2)  dynamic routing, we use a sigmoid function to replace the "routing softmax" function to strengthen discernibility of the output vector for each class and enhance the robustness capability of the CapsNet.

\subsubsection{Multi-scale parallel convolution layer}

Based on the advantages of multi-scale features in CNNs, more researchers have tried to introduce multi-scale ideas into the CapsNets model. For example, \cite{2018xiang} proposed a multi-scale capsule network, which solved the problems of the original model not well-suited to images with rich internal features. This model has shown improved the performance and convergence become easier than previous model.
\cite{2021kim} proposed a  Multi-Scale Decomposed Capsule Network (MDCN) for the issue of parameter redundancy during the training of capsule networks, this model can use fewer parameters to synthesize capsules through the MDCN architecture and has a better performance and deduced the parameters. Therefore, we have also tried to introduce features of the multiple scales in our work to improved the structure of model.

In the improved model, we increase the diversity of features and reduce the loss of basic information by combining the advantages of CNN and CapsNet models. In our work, the MSCCN model using the three convolution kernels with different scales to extract the low-level features of galaxy morphology, and excavate the multi-scale and multi-level feature information. In order to extract multi-scale morphological features of galaxy images, $7\times7$, $5\times5$ and $3\times3$ convolution kernels are selected to construct a multi-scale parallel convolution layer. 

In MSCCN model, the output of two channels is spliced to construct a complete galaxy morphological feature map. The capsule transform these feature vectors into a merged galaxy feature as the input of the following digital capsule layer.

In the primary capsule layer of the MSCCN model, the feature maps could be activated after convolution are merged, and finally a new capsule unit is synthesized per 8 channels. Therefore, the number of the capsule in the primary capsule layer is 1 / 8 of the total number of activation features in the upper layer, and the reconstructed capsule unit is used as the input of the digital capsule layer.

\subsubsection{Sigmoid Routing}
In the DigitCaps layer, the capsule network transmit information from each capsule to the next layer through the form of its activation value, and the input to each layer is a weighted summation of the activation vector $\hat{u_{i|j}}$. The process is shown in Eq.(\ref{eq:quadratic4}),
where the $c_{ij}$ is the coupling coefficients determined by dynamic routing with iterations. 

In traditional model, the $c_{ij}$ is calculated by softmax function, as in Eq.(\ref{eq:quadratic6}), but we found softmax function would convert the logits of the coupling coefficients into a set of concentrative values, it may lead to little difference in the coupling coefficients assigned to the true features and the false features. Which may resulting the wrong summation of prediction vectors in next capsule, finally affect the final performance of classification. In our study, we try using sigmoid function instead of softmax function in dynamic routing. Where the $c_{ij}$ is no longer the allocation probabilities of the capsule, but the strength of the correlation between the two capsules. It is defined by equation

\begin{equation}
    {c_{ij}^{\prime}=\frac{1}{1+exp(b_{ij})}}
    \label{eq:1}
\end{equation}

Sigmoid function is a continuous smooth function, when we used it as an activation function, which can compress the data uniformly at $(0,1)$ to enhance the expressiveness of network. In dynamic routing, sigmoid function can assign large coupling coefficients to the real features, and assign a small coupling coefficients to the false features, which could avoid wrong features after dynamic routing to obtain a larger weight coefficients to transfer to the next capsule layer. Compared with softmax, sigmoid can reduces the agglomeration effect of the coupling coefficients in the capsule and let the capsule of Digitcaps layer obtain a more uniform distribution of coupling coefficients. Table \ref{tab:sigmoid} is the update process of sigmoid dynamic routing.

\begin{table*}
	\caption{Sigmoid Routing Algorithm}
	\label{tab:sigmoid}
    \scalebox{1}{
	\begin{tabular}{l} 
		\hline
        sigmoid routing algorithm \\
        \hline
        1: Input to routing($\hat{u}_{j|i},r,l$) \\
        
        2: ~~~~For all capsule $i$ in layer $l$ and capsule $j$ in layer $(l+1)$: $b_{ij} \leftarrow 0$ \\
        
        3: ~~~~For $r$ iterations do: \\
        
        4: ~~~~~~~~~For all capsule $i$ in layer $l$: $c_{ij}^{\prime} \leftarrow sigmoid(b_{i})$ \\
        
        5: ~~~~~~~~~For all capsule $j$ in layer $l+1$: $s_{j} \leftarrow  \sum_i c_{ij}^{\prime} \cdot \hat{u}_{j|i}$\\

        6: ~~~~~~~~~For all capsule $j$ in layer $l+1$: $v_{j} \leftarrow squash(s_{j})$\\

        7: ~~~~~~~~~For all capsule $i$ in layer $l$ and capsule $j$ in layer $(l+1)$:$b_{ij} \leftarrow b_{ij}+\hat{u}_{j|i} \cdot v_{j}$\\
        
        ~~~~~~~~Return $v_{j}$\\
        \hline
	\end{tabular}}
\end{table*}

\section{Data}
\label{sec:data set}
\subsection{Data Preparation}
Our sample set were selected from the Galaxy Challenge in the Galaxy Zoo2(GZ2) (\cite{hart2016galaxy}), and the data set is deployed on the \textit{Kaggle platform}\footnote{\url{	https://www.kaggle.com/competitions/galaxy-zoo-the-galaxy-challenge/overview}}. The data of Kaggle were selected from SDSS DR7, which contains 61,578 labeled galaxies observations with a size of $424\times424\times3$ pixels, the label of each image is $1 \times 37$ vector, which comes from the cumulative frequency correction value of GZ2 volunteers voting scores. We selected five classes of galaxies from the GZ2 for model training and testing. The galaxies of five classes are completely round, in-between smooth, cigar-shape smooth, edge-on and spiral \citep{zhu2019galaxy}.

\begin{figure}
	\includegraphics[width=\columnwidth]{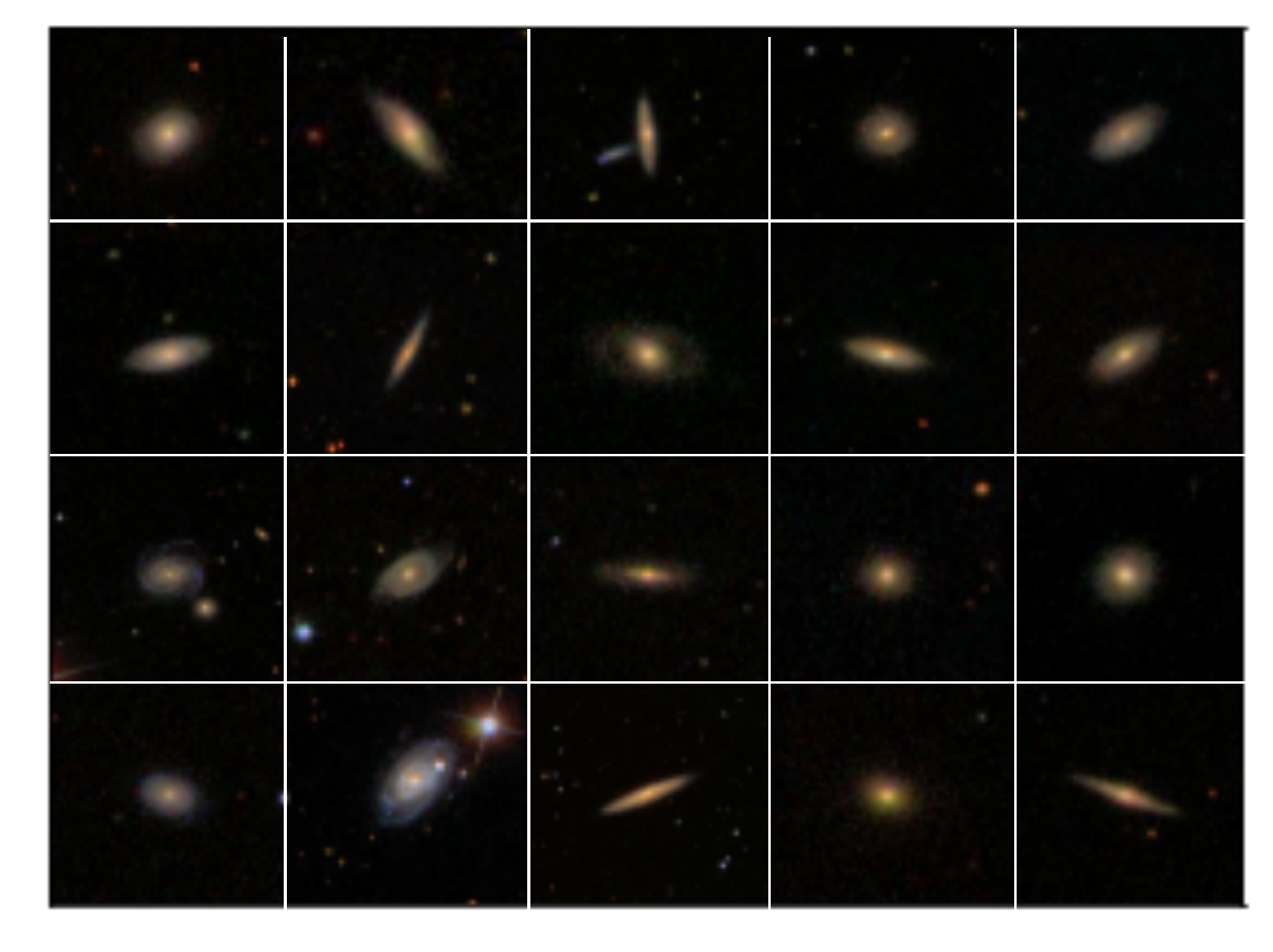}
	\caption{Galaxy morphology images in GZ2 data set}
	\label{fig:example_figure4}
\end{figure}

The data set of GZ2 has its corresponding classification threshold standard. For a galaxy image, its cumulative voting score correction value must meet a certain threshold to be classified into a galaxy category. To obtain enough sample data of galaxy, we modify the threshold selection criteria of the smooth galaxy, and the threshold criteria of other galaxy remain unchanged. Table \ref{tab:example} is the threshold selection criteria for classifying five classes galaxy.

\begin{table*}
	\centering
	\caption{The criteria of clean sample selection. Where the T01-T11 is 11 classification problems in GZ2; ${f_{smooth}}$ is the probability that a galaxy image is classified as a smooth galaxy; ${f_{features/disk}}$ is the probability of being classified as characteristic or disc structure; ${f_{spiral/yes}}$ is the probability of being classified as a spiral; ${f_{Edge-on/yes}}$, ${f_{Edge-on/no}}$ are the probability of being classified as a edge-on and not a edge-on; is the probability of being classified as a spiral; ${f_{completely - round}}$, ${f_{in - between}}$ and ${f_{cigar - shaped}}$ are the probability of being classified as completely round smooth, between a round and cigar-shape.\citep{2013MNRAS.435.2835W}}
	\label{tab:example}
    \setlength\tabcolsep{19pt}
    \scalebox{1}{
	\begin{tabular}{cccrl} 
		\hline
		Class & Clean sample & Task & Threshold & ${N_{sample}}$\\
		\hline
		\multirow{3}{*}{0} & \multirow{3}{*}{Spiral} & T01 & ${f_{features/disk}} \ge 0.430$ & \multirow{3}{1cm}{7806}\\
		  &  & T02 & ${f_{edge - on,no}} \ge 0.715$ & \\
		  &  & T04& ${f_{spiral,yes}} \ge 0.619$ & \\
		\hline
		\multirow{2}{*}{1} & \multirow{2}{*}{Edge-on} & T01& ${f_{features/disk}} \ge 0.430$ &\multirow{2}{1cm}{3903} \\
		 &  & T02 & ${f_{edge - on,yes}} \ge 0.602$ & \\
		\hline
		\multirow{2}{*}{2} & Cigar- & T07 & ${f_{smooth}} \ge 0.469$ & \multirow{2}{1cm}{578}\\
		 & shape smooth & T01 & ${f_{cigar - shaped}} \ge 0.50$ & \\
		\hline
		 \multirow{2}{*}{3}& Completely  & T07 & ${f_{smooth}} \ge 0.469$ & \multirow{2}{1cm}{8343}\\
		 & round & T01& ${f_{completely - round}} \ge 0.50$ & \\
		\hline
		\multirow{3}{*}{4} & In-between  & T07 & ${f_{smooth}} \ge 0.469$ & \multirow{3}{1cm}{8069}\\
		 & smooth & T01 & ${f_{in - between}} \ge 0.50$ & \\
		\hline
	\end{tabular}}
\end{table*}

We collected 28790 galaxy images based on the threshold rule for galaxies in Table \ref{tab:example}. Among them, 7806 spiral galaxies, 578 cigar galaxies, 3903 lateral galaxies, 8069 intermediate galaxies and 8434 circular galaxies. Finally, a data set had been constructed in our study.

\subsection{Data Preprocessing}
When train set is limited, data augmentation can improve the performance of the model. Data augmentation plays an important role in the final recognition performance and generalization ability of a model. General data augmentation methods include rotation, translation, scaling, random flipping, and brightness. It should be noted that scaling, translation and brightness change had little effect on the model performance. Therefore, we augment the sample set through rotation and random flipping(horizontal and vertical).
However, excessive data augmentation may lead to increased the computational effort of model. Sometimes, although resulting in overall performance improvement, imbalance problem between classes may occur in it \citep{2022arXiv220403632B}. Therefore, 
we experimented with augmenting the train set by a factor of 1 to 5, and found that the accuracy of the model did not change significantly when the training set was augmented by a factor of 3.
 Table \ref{tab:landscape} shows that the aggregation of the model accuracy and training time at different augmentation multiples.

  \begin{table*}
  \caption{Model performance and training time under different data augmentation methods The "Multiples" is augmentation multiples of train set.}
  \label{tab:landscape}
  \setlength\tabcolsep{19pt}
  \begin{tabular}{ccccc}
    \hline
    \multirow{2}{*}{Multiples} & \multicolumn{2}{c}{Data Augmentation Methods}& \multirow{2}{*}{Accuracy} & \multirow{2}{*}{Time($s/Epoch$)} \\
     & Rotation  Angle($^{o}$) & Flipping Probability&  &  \\
    
    \hline
    0 &   /  &   /   &   0.9505  & 16 \\
    1 &  90  &   /   &0.9659  & 30 \\
    2 &  90  &   0.5   & $\bm{0.9701}$  & 46 \\
    3 & 90,180  &  0.5    &0.9698  & 65 \\
    4 & 90,180,270  &   0.5   &0.9669  & 89 \\
    5 & 45,90,180,270  &   0.5   &0.9689  & 145 \\
    \hline
  \end{tabular}
 \end{table*}

Secondly, we perform intermediate cropping of galaxies images to retain the complete information of galaxy images, to reduce noise and dimension of data. 
We cut the images from 424 $\times$ 424 $\times$ 3 pixels to 212 $\times$ 212 $\times$ 3 pixels, and then down-sample it to 80 $\times$ 80 $\times$ 3 pixels. After preprocessing the data, we divide the data into test set and training set at a ratio of 1:9.

\section{Result and Discussion}\label{sec:results}

\subsection{Computer Setup}
The primary device for this study is a server with a 24G RTX A5000 GPU, 14-core Intel (R) Xeon (R) Gold 6330 CPU, besides Windows10 OS, 2021.1 version of Pycharm professional version, 11.2 version CUDA, Python language, Tensorflow, Pandas, and Scikit-learn libraries.

\subsection{Selection of hyper-parameters}
In deep learning, hyper-parameters selection plays a vital role in performance of model. We carried out a series of experiments on the hyper-parameters such as convolution kernel size, dynamic routing number and batch-size to select the best value of hyper-parameters.

The convolution layer of the capsule network is mainly used to extract the low-level features of the galaxy images, and the size of the convolution kernel affects the performance of feature extraction. We designed three experiments based on the principle that large convolution kernels can expand the acceptance domain and small convolution kernels can extract more detailed features. The selected convolution kernels are 7$\times$7, 5$\times$5 and 3$\times$3 combinations, 9$\times$9, 6$\times$6 and 3$\times$3 combinations, and the single layer convolution with a kernel size of 9. Other parameters of network remain unchanged and the model is trained in batches. The results are shown in Table \ref{tab:exampl}.

The weight updating in CapsNet is based on iterative dynamic routing, and the number of dynamic routing plays an essential role in the stability and classification performance of the model. To explore the influence of dynamic routing numbers on the performance of CapsNet, we take the dynamic routing numbers of 3, 6, and 12, and other parameters of network remain unchanged. The classification results are shown in Table \ref{tab:exampl}.

Deep learning is optimized by a stochastic gradient descent algorithm. The principle of the stochastic gradient descent algorithm is as follows:
\begin{equation}
	{w_{t + 1}} = {w_t} - \eta \frac{1}{n}\sum\limits_{x \in \beta } {\Delta l(x,{w_t})} 
	\label{eq:quadratic}
\end{equation}
where the $\eta $ is learning rate, $\mathit{n}$ is batch-size, $w_t$ is the gradient, in Eq.(\ref{eq:quadratic}), the two parameters directly determine the optimization performance of the model, those are the most critical parameters for the optimization convergence of the model. In our work, the learning rate as 0.001 and decayed over time. the several experiments was conducted on the batch-size to select the best value. In this study, the batch-size is 128, 64, 32, and other parameters were unchanged, the classification results are shown in Table \ref{tab:exampl}.

\begin{table}
	\caption{The classification accuracy of MSCCN under different hyper-parameters.}
	\label{tab:exampl}
	\setlength\tabcolsep{22pt}
	\begin{tabular}{ccc} 
		\hline
		\multicolumn{2}{|c|}{Hyper-parameters}  & Accuracy\\
		\hline
		\multirow{3}*{Conv Kernel} & 7$-$5,3 & $\bm{0.9673}$\\
		~ & 9$-$6,3 & 0.9396\\
		~ & 9 & 0.9293\\
		\hline
		\multirow{3}*{Routing} & 3 & 0.9273\\
		~ & 6 & 0.9144\\
		~ & 12 & $\bm{0.9664}$\\
		\hline
		\multirow{3}*{Batch-size} & 32 & $\bm{0.9626}$\\
		~ & 64 & 0.9466\\
		~ & 128 & 0.9473\\
		\hline
	\end{tabular}
\end{table}

\subsection{Regression Model Based on MSCCN}

In our work, we also designed a regression model based on the CapsNet, which predicted the vote fraction probabilities of 37 questions list in the GZ2 decision tree. In this regression model, we had not use data augmentation and only crop the images with a central window like our classification model. The final input of regression is an image of size $224\times224\times3$ and vote-fractions for the 37 questions in the GZ2 decision tree.

Root mean squared error (RMSE) is to measure the deviation between the predicted value and the actual values. The RMSE of our regression model is 0.08192, when comparing our RMSE with the public leader-board of Kaggle Galaxy Zoo challenge, we find the RMSE were placed ninth on the public leader-board. The results show the error between the predicted value and the actual values is very slight, our model can correctly predict the probabilities.

\subsection{Classification Results of the MSCCN model}
When the convolution kernel size is divided into 7 $\times$ 7, 5 $\times$ 5 and 3 $\times$ 3, batch-size is 32, and the number of dynamic routing is 3, the MSCCN model will be achieved the best results. We set the initial learning rate as 0.001 and the learning rate decay factor as 0.9. As the iteration increases, the learning rate decays in the proportion of 0.9 to avoid over-fitting of the classification model. There are five classes of the galaxy in data set, therefore we set the number of digital capsules as 5.

\begin{figure}
	\includegraphics[width=\columnwidth]{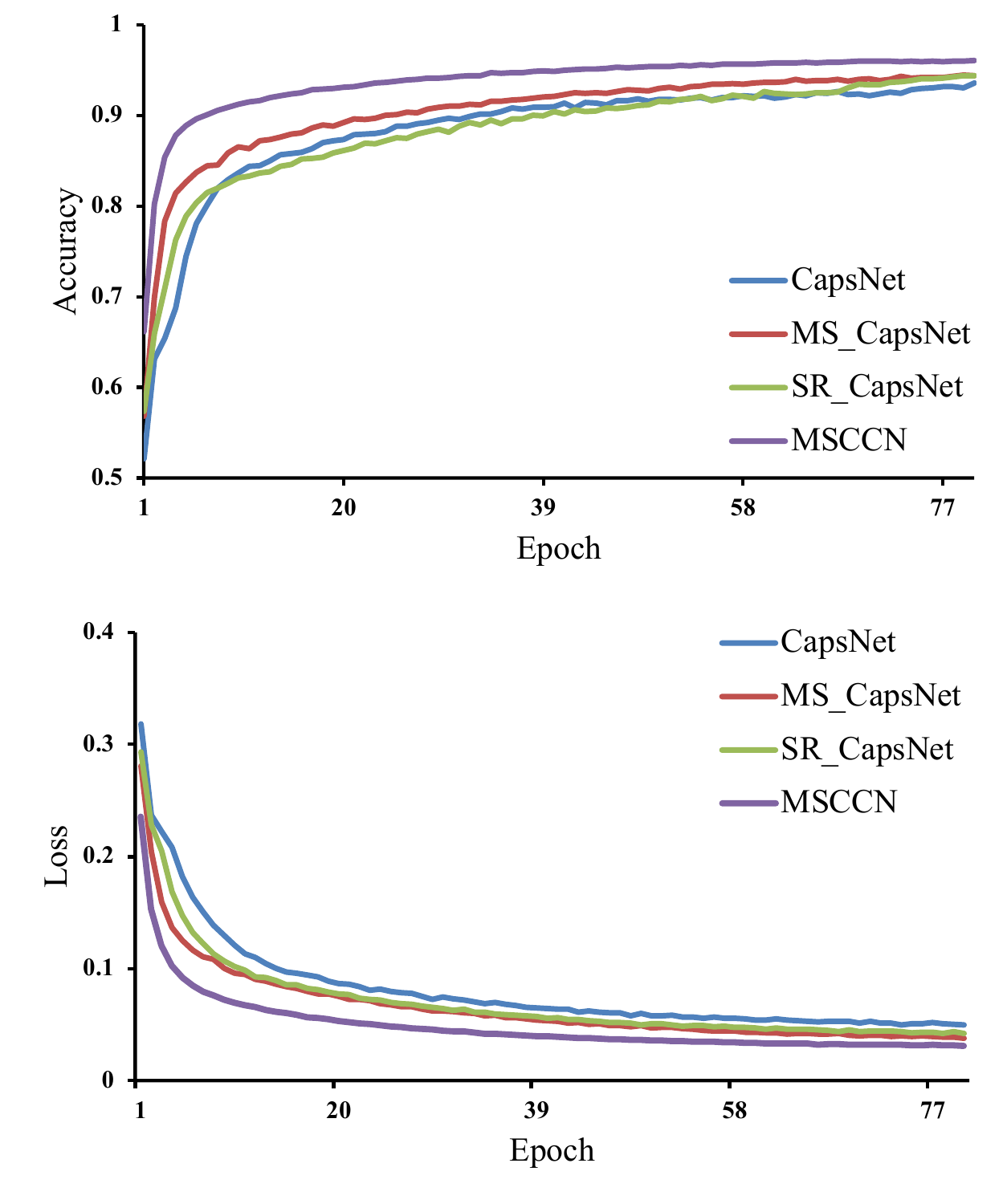}
	\caption{The accuracy and loss curve of the different model. The CapsNet is the traditional models, the MS\_CapsNet is the CapsNet model with a multi-scale convolution layer, the SR\_CapsNet is the CapsNet model with a sigmoid routing.}
	\label{fig:example_figure5}
\end{figure}

The final classification accuracy of the well-trained model for galaxy morphology is $0.9701$. Among the five classes of galaxy morphology classification results, the edge-on galaxy obtained the highest classification accuracy of $0.9925$, the second is completely round smooth galaxy, its accuracy is $0.9751$. In addition, the classification accuracy of spiral galaxies and in-between smooth galaxies are $0.9725$ and $0.9617$. The classification accuracy of cigar-like galaxies is $0.9298$.
The analysis results shows that the original data of the cigar-shape smooth galaxy is fewer, therefore the MSCCN model cannot fully learn the characteristics of cigar shape smooth galaxy, and the classification performance is lower than the other four classes. Fig.\ref{fig:example_figure5} is the curve of training loss and classification accuracy of the MSCCN model. In our model, the model converges after 70 epochs, and the number of parameters are 7.65 Million. Training an epoch takes 46s, and a well-trained model will spent 53 minutes.

\begin{figure}
	\includegraphics[width=\columnwidth]{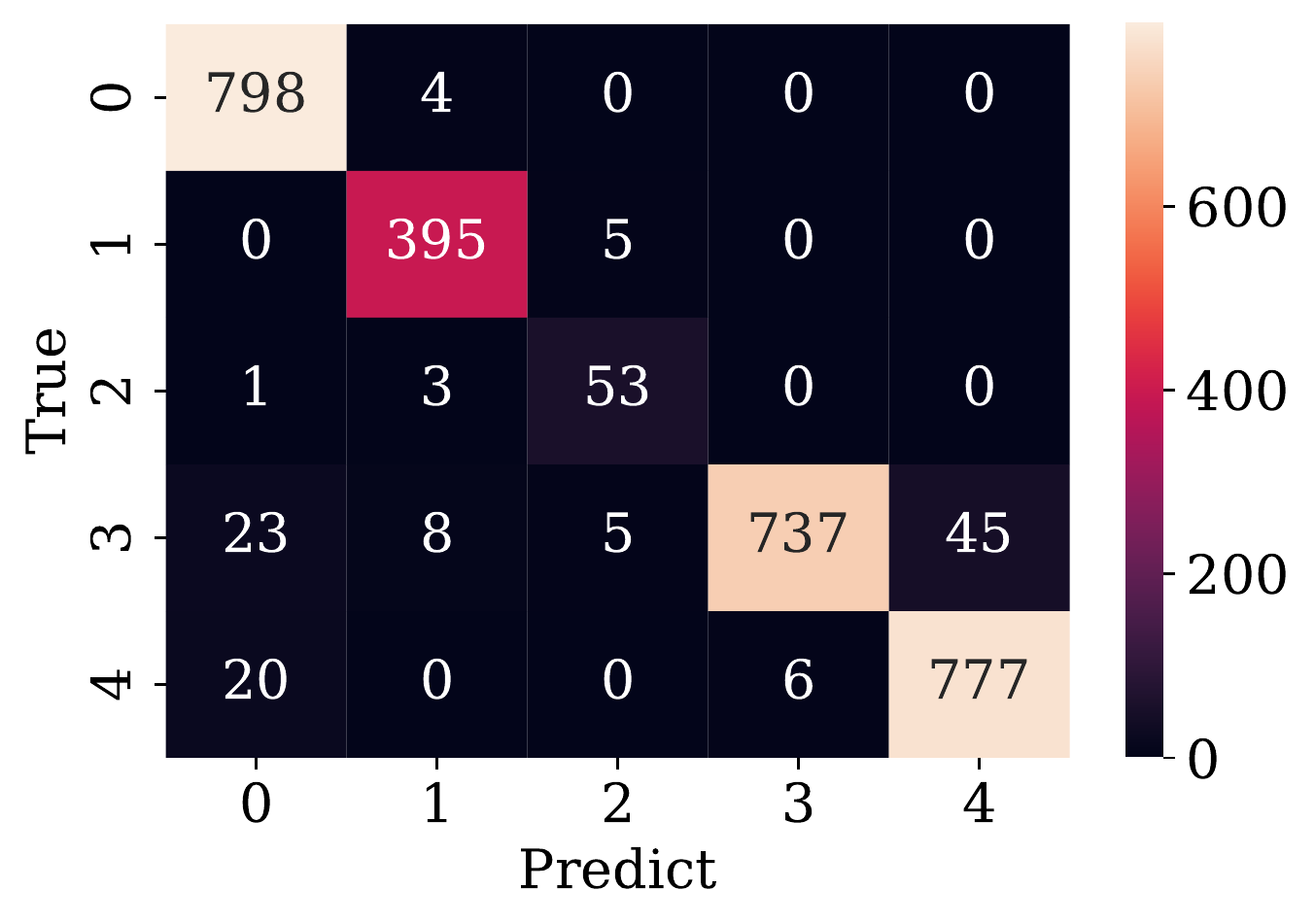}
	\caption{The confusion matrix of MSCCN model}
	\label{fig:example_figure6}
\end{figure}

At the same time, we calculated the confusion matrix of the testing set for the MSCCN model. In Fig.\ref{fig:example_figure6}, each row of the matrix represent true label of the galaxy category, while each column represent the predicted label of the galaxy category. From the confusion matrix, we found that there are 3 edge-on galaxies wrongly predicted as cigar-shape smooth galaxies, and 1 cigar-shape smooth wrongly classified as spiral galaxies, because spiral galaxies and cigar-shape smooth galaxies are very similar in morphology and structure. After analysis of the study results, we find that when the characteristics of the category are similar or same, the performance of model will be inhibited. In addition, 6 completely round smooth galaxies were misclassified as in-between smooth galaxies, and 45 in-between smooth galaxies were predicted to be completely round smooth galaxies. Our analysis shows that the shapes of these two galaxies are smooth and the threshold selection of clean samples between them is very close, which results in some deviations in the classification results.

\begin{figure}
	\includegraphics[width=\columnwidth]{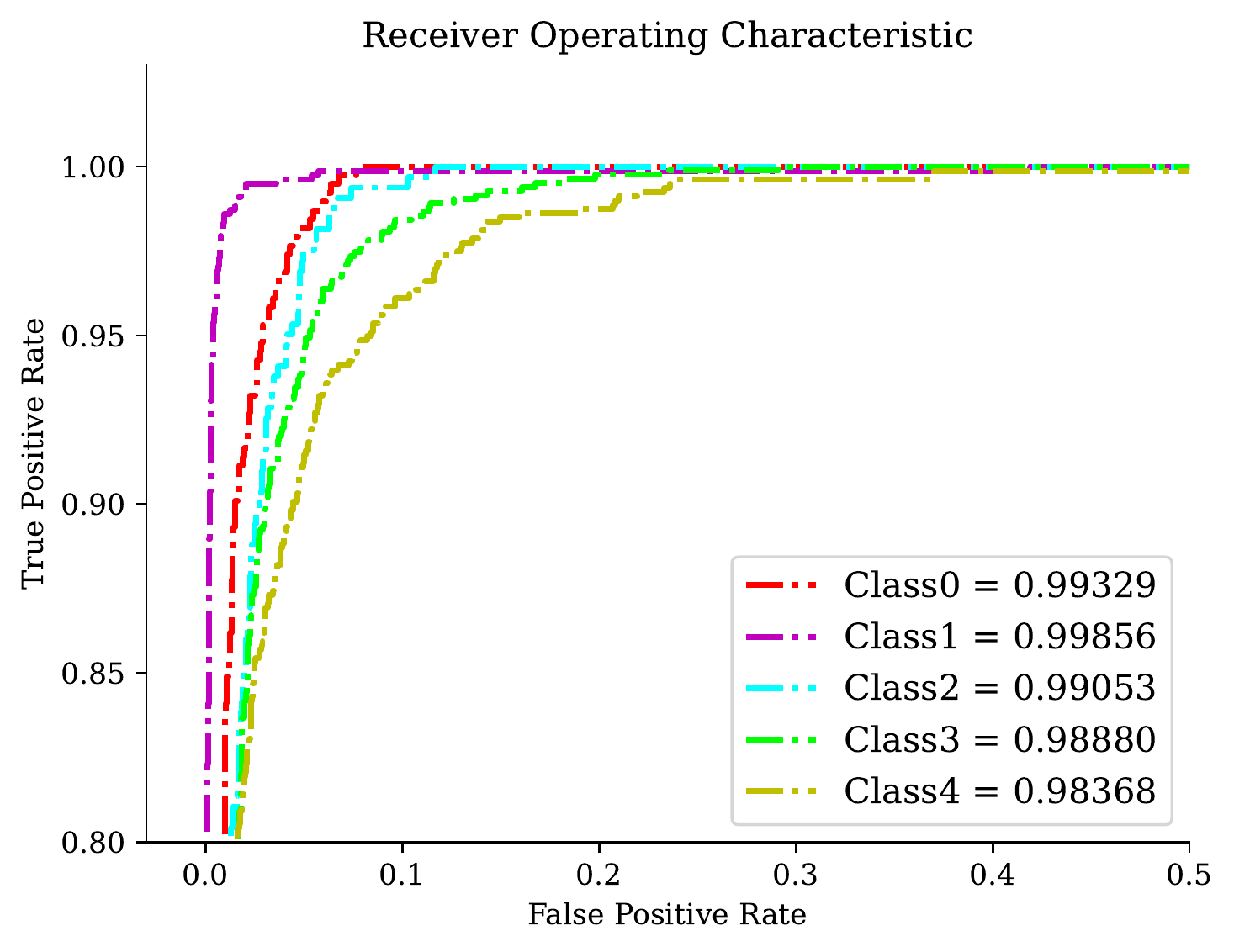}
	\caption{ROC curves and AUC values of MSCCN model}
	\label{fig:figure7}
\end{figure}

Receiver Operating Characteristic curve (ROC) and Area under the curve (AUC) can reduce the interference caused by different test sets in model evaluation, they can more objectively measure the performance of model. To verify the generalization ability of model, we calculated the ROC curve and the AUC value of the model for each classes of galaxy. 
In Fig.\ref{fig:figure7}, each colors represents a category of the galaxy. The horizontal axis is  the False Positive Rate (FPR), and the vertical axis is the True Positive Rate (TPR). The TPR of ideal model is supposed to be close to 1, and FPR close to 0. In Fig.\ref{fig:figure7}, the TPR of the five galaxy classes is close to 1, and the FPR is close to 0, indicating that the model has achieved good predicted results for each galaxy classes. The AUC values of the former four types of galaxies are all above 0.99. And the AUC values of the in-between smooth galaxies are also above 0.98, which indicates that the robustness of the model is relatively strong, and the imbalance of data samples has little effect on the overall performance of the model.

\subsection{Results Comparison with Other Similar Works}

\citet{dieleman2015rotation} first applied deep learning to the galaxy morphology classification in the galaxy challenge. They proposed a convolution neural network galaxy morphology classification model specifically for galaxy image properties. The model effectively uses the translation and rotation symmetry in the image and autonomously learns the abstract representation of multi-level features of the image. It can efficiently and automatically represent image categories with morphological information and classify galaxies accurately and quickly. \citet{gravet2015catalog} used the Dieleman model to classify the high-redshift galaxy images from CANDELS cruise data into five categories and achieved excellent classification results. In 2018, \citet{zhu2019galaxy} based on ResNet V2 and combined with the characteristics of galaxy images themselves, proposed an improved deep residual network model for galaxy morphology classification, namely ResNet-26. This model improves the residual unit, while reducing the depth of network, widening the width of network, and realizing the automatic extraction of galaxy morphological features for identification and classification. 
In 2022, \citet{Gupta2022} introduced a continuous depth version of the Residual Network (ResNet) called Neural ordinary differential equations (NODE) for galaxy morphology classification. They train NODE with different numerical techniques such as Adjoint and Adaptive Checkpoint Adjoint (ACA) and compared them with ResNet, the results show that the accuracy of NODE is comparable to ResNet, and the number of parameters used is about 1/3 compared to ResNet.

In this work, we compared the MSCCN model with Dieleman model, Resnet-26 model, NODE model in the same experimental environment, to verify the validity and superiority of the MSCCN model in the task of galaxy morphology classification. The Dieleman model is the first model to apply CNN to astronomical image classification, which consists of 4 convolution layers and 3 full connection layers. The Resnet-26 model is a 26-layers residual network. We analyzed the results of the three models and selected the accuracy, precision, recall, and F1-score as the evaluation indexes of the model. Table \ref{tab:exam} shows the study results of four models.

\begin{table*}
	\centering
	\caption{Comparison results of classification evaluation indexes of four models.}
	\label{tab:exam}
	\setlength\tabcolsep{2.5pt}
	\scalebox{1}{
	\begin{tabular}{ccccc} 
		\hline
		Model & Dieleman(\cite{dieleman2015rotation}) &NODE(\cite{Gupta2022}) & Resnet-26(\cite{zhu2019galaxy}) & MSCCN(In this work) \\
		\hline
		Accuracy($\%$)  & 93.88 & 91.65 & 94.68 & $\bm{97.01}$\\
		Precision($\%$)  & 94.55 & 91.55 & 95.12 & $\bm{95.97}$\\
        Recall($\%$)  & 94.86 & 93.59 & 95.21 & $\bm{98.16}$\\
		F1 ($\%$) & 94.56 & 92.60 & 95.15 & $\bm{96.39}$\\
		\hline
	\end{tabular}}
\end{table*}

Dieleman, NODE model, and RESNET-26 model are all deep learning models designed for the task of galaxy morphological classification. From Table \ref{tab:exam}, it shows that they have achieved good effects in the classification of galaxy images, the accuracy and precision of those methods are 0.9388, 0.9165, 0.9468 and 0.9455, 0.9155, 0.9512. The classification accuracy and precision of MSCCN model are $0.9701$ and $0.9597$, which is better than the other methods. We introduced Recall as the same time. Dieleman, NODE, and Resnet-26 are 0.9486, 0.9359 and 0.9521 on the recall of the model, and the Recall of MSCCN model is $0.9815$. The results shows when the number of Cigar-shape smooth galaxy samples is fewer, and the data set samples are unbalanced, the MSCCN model still performs better than the former three groups of models in the comparative test. The F1-score of the MSCCN is also better than those of the other three models, and the F1 scores of the four models are $0.9639$, 0.9260, 0.9515, 0.9456.

\citet{2020arXiv200813611K} proposed a fine-tuned architecture using EfficientNetB5 to classify galaxies into 7 classes. They introducing irregular galaxies on top of the 5 classes, and subdividing spiral galaxies into barred spiral galaxies and unbarred spiral galaxies based on whether they have a bar structure. The fine-tuned architecture achieved a classification accuracy of 0.9370. \citet{2021arXiv211001024Y} used a Vision Transformer to classify the smaller-sized and fainter galaxies. In their work, they classified irregular galaxies and merger galaxies based on the work of \citet{zhu2019galaxy} and \citet{2020arXiv200813611K} according to the characteristics of whether they have mergers or not, and then classified galaxies into 8 classes, the best overall classification accuracy of this work is 0.8055. To evaluate our model more objectively, we select 8 classes of galaxies according to \citet{2021arXiv211001024Y}, and select 7 classes of galaxies based on \citet{2020arXiv200813611K}. When we use MSCCN to classify the 8 classes of  galaxies, the accuracy is 0.9159, when the galaxies are of 7 classes, the accuracy is 0.9427. The results shows that MSCCN model still perform well on multiple classes of galaxy morphology classification task.

\subsection{Analysis and Discussion}
MSCCN model shows excellent results in accuracy, precision, loss, confusion matrix, ROC curve and other model evaluation indices, which shows that the model has a good performance for the images of galaxy data. The accuracy of the cigar-shape smooth galaxy is 0.9298 (Without data augment is 0.7236), which is lower than the other four classes. Which is due to the number of original samples of the cigar-shape smooth galaxies is too few. And when the model classifies galaxies, the classification boundary tends to occupy the area of minority classes. However, through the analysis of the ROC curve and AUC value of the model, we found that the data imbalance has a limited impact on the overall performance of the model, and the overall generalization ability of the model is still strong. In data set, the completely round smooth galaxy and the in-between smooth galaxy have the characteristics of none obvious interclass boundary and little difference in classification threshold, which undoubtedly increases the difficulty of classification. The classification accuracy of the model for completely round smooth galaxies and in-between smooth galaxies is 0.9706 and 0.9810. The results of this work show that the multi-scale convolution layer of the MSCCN model can extract the multi-scale primary features of the galaxy, accurately classify each classes of galaxy, and eliminate the influence of sample classification boundary ambiguity on the performance of the model.

\section{Visualization Analysis of MSCCN Model}
\label{sec:Visualization}
\begin{figure}
	\includegraphics[width=\columnwidth]{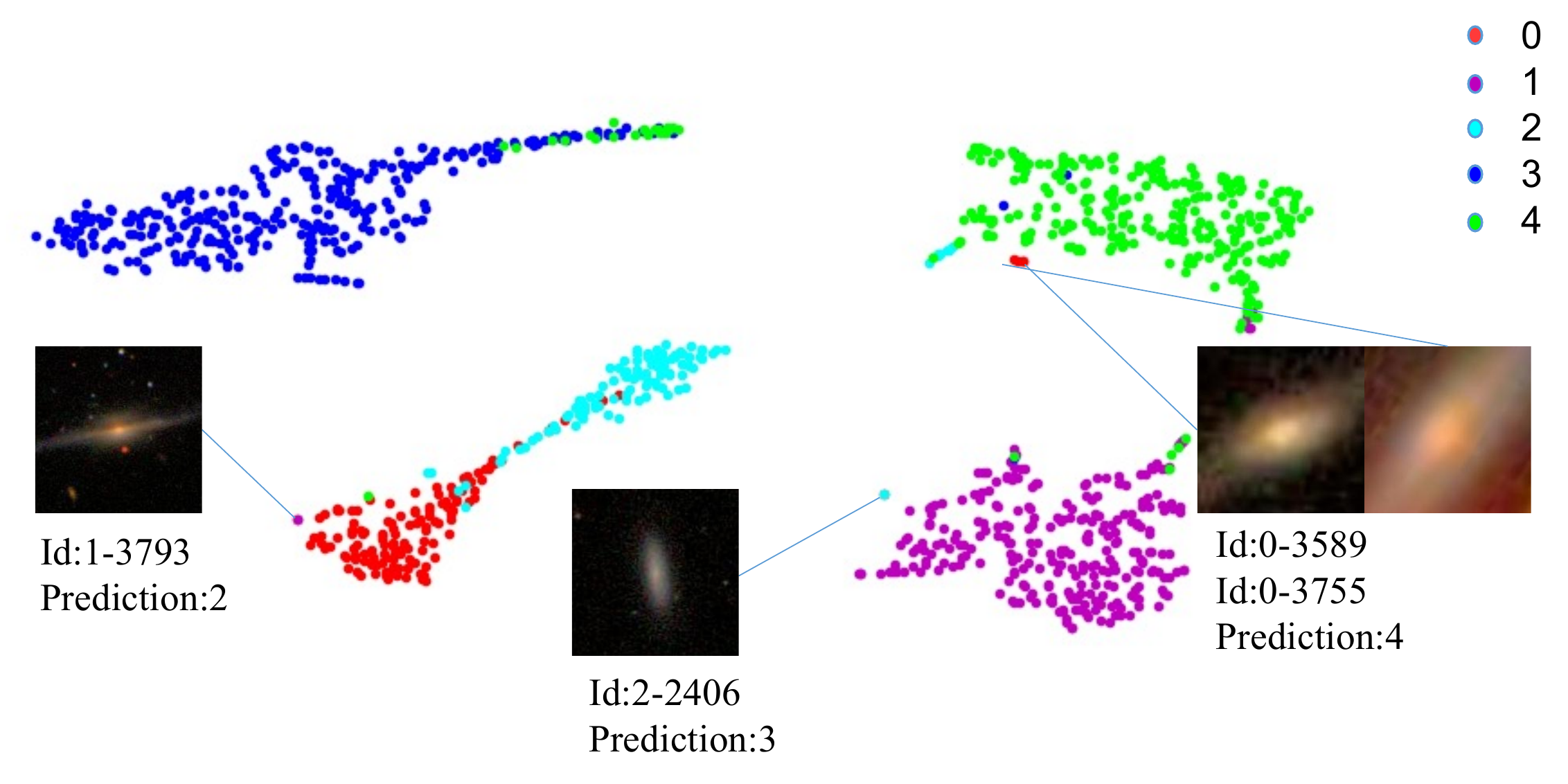}
	\caption{Feature visualization of DigitCaps layer of MSCCN model}
	\label{fig:figures}
\end{figure}

In order to explore the information on the representation of the morphological features of the galaxy from the data itself, we randomly selected 1000 samples from the test set for visual representation to analyze the output of the DigitCaps layer of the MSCCN model. The visualization analysis was implemented in our study based on the t-SNE algorithm, that is a nonlinear dimensionality reduction and visualization method\citep{van2008visualizing}. It can retain the local structure of the sample data and obtain low-dimensional data with higher similarity to the original high-dimensional data \citep{2018arXiv180705657D}. The t-SNE algorithm converts the similarity between data points into probability, and the similarity in the original high-dimensional space is represented by Gaussian distribution. The probability of embedding space is represented by T-distribution so that the data in the high-dimensional space is mapped to the low-dimensional space and visual representation. Fig.\ref{fig:figures} is the feature visualization the MSCCN model.

In Fig.\ref{fig:figures}, because the galaxies of the same morphology have similar underlying structures, each class of galaxies is distributed in clusters. Completely round smooth galaxy and in-between smooth galaxy tend to converge. Since both completely round smooth galaxy and in-between smooth galaxy are smooth galaxy, there is no definite classification boundary between them, and they are similar in shape, resulting in misclassified. In Fig.\ref{fig:figures}, cigar-shape smooth galaxies and spiral galaxies are inter-weaved, and many samples between them are misclassified. Through analyzing the results and comparing  the image, we found that the geometric shapes of cigar-shape smooth galaxy and lateral spiral galaxy are very similar. When labeling the original samples, the labels of the two galaxies are easily mislabeled, but they are correctly recognized by the model when classified. At the same time, the shapes of the two galaxies are similar, which can also lead to the wrong classification. This discovery contribute to the understanding of  the physical properties of galaxy morphology.

\begin{figure}
	\includegraphics[width=\columnwidth]{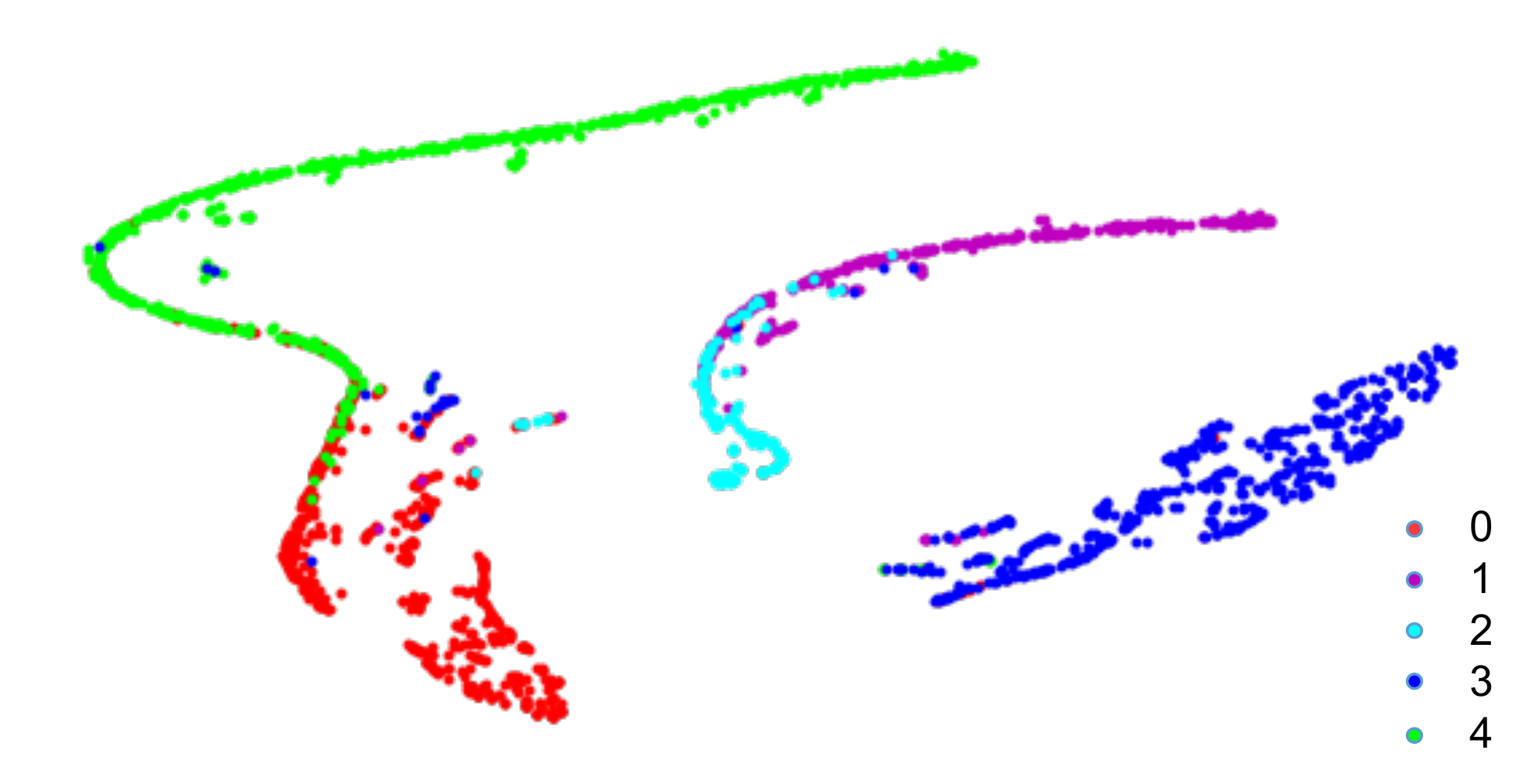}
	\caption{Feature visualization of last average pooling layer of CNN (Accuracy $=0.92$)}
	\label{fig:figure9}
\end{figure}

Fig.\ref{fig:figures} shows the visual representation of outliers in the DigitCaps layer of MSCCN model. The red outliers (Id: 1-3589, 1-3755) belong to spiral galaxies but appear in the in-between smooth galaxy represented by orange. The analysis suggests that the two galaxy images have similar structures to the in-between smooth galaxy, therefore it is early been misclassified. The model misclassified the blue outliers (Id: 3-2476) as spiral galaxies. After analysis, it is a galaxy image with poor quality. The galaxies are located in a small part of the image, and it is not easy to distinguish the class of the galaxies. In addition, the outliers (Id: 2-3793) were found to be a spiral galaxy, but it was wrongly marked as an in-between smooth galaxy. It was predicted as a lateral galaxy after the model classification.

Furthermore, we visualized the output of the last average pooling layer of the CNN as a comparison to our visualization work. In this part work, we constructed a convolution neural network to classify galaxy morphology based on the model of \citep{2018arXiv180705657D}. Finally, we select 1000 samples to visualized the features of CNN, and Fig.\ref{fig:figure9} is the feature visualization of last average pooling layer of CNN. We found our MSCCN model has a better separability compared to CNN, and each class of galaxy is separated from the other.

We visually analyze the output of the DigitCaps layer. In Fig. \ref{fig:figures}, each class of galaxy is clustered and separated from each other, which shows that the MSCCN model has better separability, and the classification effect of the model is excellent. In the future, we try to explore the low-level structure of sample data and the physical meaning of high-dimensional feature representation through visualization analysis, which is contribute to discover data rules quickly, explain the classification results of model, and provide more helpful feedback for the galaxies classification system.

\section{Conclusion}
\label{sec:conclusion}
This study present a method for galaxy morphology classification, namely multi-scale convolution capsule network. In this network, we used a multi-scale convolution layer to replace the convolution layer of the traditional capsule network, built a parallel convolution layer and reconstruct the main capsule layer. Additionally, we used a disperse dynamic routing algorithm to get a more uniform distribution of coupling coefficients, it can assign a larger coupling coefficients to the true features and smaller coupling coefficients to the wrong features, which can strengthen the discernibility of the output vector for each class and improve the robustness of the CapsNet. The model can fully capture the multi-scale galaxy features, further extract the hidden galaxy information in the galaxies image, and reduce the parameter redundancy. At the same time, it solved the problem that traditional deep learning cannot extract the spatial information of the galaxy and the loss of feature information. The results show that the multi-scale convolution capsule network has a better classification performance on galaxy morphology classification. The classification results are better than the comparative models selected in this study and can be applied to galaxy morphology classification.

\section{Acknowledgements}

We thank the anonymous referee for valuable and helpful comments and suggestions. This work is supported by the National Nature Science Foundation of China (61561053), and the Scientific Research Foundation Project of Yunnan Education Department(2023J0624). This work is also supported by the Astronomical Big Data Joint Research Center, co- founded by National Astronomical Observatories, Chinese Academy of Sciences and Alibaba Cloud.

\section*{Data Availability}
All data used in this work are publicly available. Details on how to access the data can be found on their websites: \url{https://www.kaggle.com/competitions/galaxy-zoo-the-galaxy-challenge/overview}. 

And all code produced in this work are available upon reasonable request to the authors.



\bibliographystyle{mnras}
\bibliography{example} 

\begin{thebibliography}{}
\makeatletter
\relax
\def\mn@urlcharsother{\let\do\@makeother \do\$\do\&\do\#\do\^\do\_\do\%\do\~}
\def\mn@doi{\begingroup\mn@urlcharsother \@ifnextchar [ {\mn@doi@}
  {\mn@doi@[]}}
\def\mn@doi@[#1]#2{\def\@tempa{#1}\ifx\@tempa\@empty \href
  {http://dx.doi.org/#2} {doi:#2}\else \href {http://dx.doi.org/#2} {#1}\fi
  \endgroup}
\def\mn@eprint#1#2{\mn@eprint@#1:#2::\@nil}
\def\mn@eprint@arXiv#1{\href {http://arxiv.org/abs/#1} {{\tt arXiv:#1}}}
\def\mn@eprint@dblp#1{\href {http://dblp.uni-trier.de/rec/bibtex/#1.xml}
  {dblp:#1}}
\def\mn@eprint@#1:#2:#3:#4\@nil{\def\@tempa {#1}\def\@tempb {#2}\def\@tempc
  {#3}\ifx \@tempc \@empty \let \@tempc \@tempb \let \@tempb \@tempa \fi \ifx
  \@tempb \@empty \def\@tempb {arXiv}\fi \@ifundefined
  {mn@eprint@\@tempb}{\@tempb:\@tempc}{\expandafter \expandafter \csname
  mn@eprint@\@tempb\endcsname \expandafter{\@tempc}}}

\bibitem[\protect\citeauthoryear{{Abraham}, {van den Bergh}  \&
  {Nair}}{{Abraham} et~al.}{2003}]{2003ApJ...588..218A}
{Abraham} R.~G.,  {van den Bergh} S.,   {Nair} P.,  2003, \mn@doi [\apj]
  {10.1086/373919}, \href
  {https://ui.adsabs.harvard.edu/abs/2003ApJ...588..218A} {588, 218}

\bibitem[\protect\citeauthoryear{{Balestriero}, {Bottou}  \&
  {LeCun}}{{Balestriero} et~al.}{2022}]{2022arXiv220403632B}
{Balestriero} R.,  {Bottou} L.,   {LeCun} Y.,  2022, arXiv e-prints, \href
  {https://ui.adsabs.harvard.edu/abs/2022arXiv220403632B} {p. arXiv:2204.03632}

\bibitem[\protect\citeauthoryear{Ball, Loveday  \& Brunner}{Ball
  et~al.}{2008}]{8193490}
Ball N.~M.,  Loveday J.,   Brunner R.~J.,  2008, \mn@doi [MNRAS]
  {10.1111/j.1365-2966.2007.12627.x}, 383, 907

\bibitem[\protect\citeauthoryear{Bershady, Jangren  \& Conselice}{Bershady
  et~al.}{2000}]{bershady2000structural}
Bershady M.~A.,  Jangren A.,   Conselice C.~J.,  2000, \mn@doi [AJ]
  {10.1086/301386}, 119, 26,45

\bibitem[\protect\citeauthoryear{{Conselice}}{{Conselice}}{2003}]{2003ApJS..147....1C}
{Conselice} C.~J.,  2003, \mn@doi [\apjs] {10.1086/375001}, \href
  {https://ui.adsabs.harvard.edu/abs/2003ApJS..147....1C} {147, 1}

\bibitem[\protect\citeauthoryear{{Dai} \& {Tong}}{{Dai} \&
  {Tong}}{2018}]{2018arXiv180705657D}
{Dai} J.-M.,  {Tong} J.,  2018, arXiv e-prints, \href
  {https://ui.adsabs.harvard.edu/abs/2018arXiv180705657D} {p. arXiv:1807.05657}

\bibitem[\protect\citeauthoryear{Dieleman, Willett  \& Dambre}{Dieleman
  et~al.}{2015}]{dieleman2015rotation}
Dieleman S.,  Willett K.~W.,   Dambre J.,  2015, \mn@doi [MNRAS]
  {10.1093/mnras/stv632}, 450, 1441

\bibitem[\protect\citeauthoryear{{Fanson} \& {Fazio}}{{Fanson} \&
  {Fazio}}{1998}]{1998SPIE.3356..478F}
{Fanson} J.~L.,  {Fazio} 1998, in {Bely} P.~Y.,  {Breckinridge} J.~B.,  eds,
  Vol. 3356, SPIE. pp 478--491, \mn@doi{10.1117/12.324471}

\bibitem[\protect\citeauthoryear{Fielding, Nyirenda  \& Vaccari}{Fielding
  et~al.}{2022}]{fielding2022classification}
Fielding E.,  Nyirenda C.~N.,   Vaccari M.,  2022, in 2022 International
  Conference on Electrical, Computer and Energy Technologies (ICECET). pp~1--6

\bibitem[\protect\citeauthoryear{Gardner et~al.,}{Gardner
  et~al.}{2006}]{gardner2006james}
Gardner J.~P.,  et~al., 2006, \mn@doi [Space Science Reviews]
  {10.1007/s11214-006-8315-7}, 123, 485

\bibitem[\protect\citeauthoryear{{Gupta}, {Srijith}  \& {Desai}}{{Gupta}
  et~al.}{2022}]{Gupta2022}
{Gupta} R.,  {Srijith} P.~K.,   {Desai} S.,  2022, \mn@doi [Astronomy and
  Computing] {10.1016/j.ascom.2021.100543}, \href
  {https://ui.adsabs.harvard.edu/abs/2022A&C....3800543G} {38, 100543}

\bibitem[\protect\citeauthoryear{Hart et~al.,}{Hart
  et~al.}{2016}]{hart2016galaxy}
Hart R.~E.,  et~al., 2016, \mn@doi [MNRAS] {10.1093/mnras/stw1588}, 461, 3663

\bibitem[\protect\citeauthoryear{{Hubble}}{{Hubble}}{1926}]{1926ApJ....64..321H}
{Hubble} E.~P.,  1926, \mn@doi [\apj] {10.1086/143018}, \href
  {https://ui.adsabs.harvard.edu/abs/1926ApJ....64..321H} {64, 321}

\bibitem[\protect\citeauthoryear{{Huertas-Company} et~al.,}{{Huertas-Company}
  et~al.}{2015}]{gravet2015catalog}
{Huertas-Company} M.,  et~al., 2015, \mn@doi [\apjs]
  {10.1088/0067-0049/221/1/8}, \href
  {https://ui.adsabs.harvard.edu/abs/2015ApJS..221....8H} {221, 8}

\bibitem[\protect\citeauthoryear{{Ivezi{\'c}} et~al.,}{{Ivezi{\'c}}
  et~al.}{2019}]{2019ApJ...873..111I}
{Ivezi{\'c}} {\v{Z}}.,  et~al., 2019, \mn@doi [\apj]
  {10.3847/1538-4357/ab042c}, \href
  {https://ui.adsabs.harvard.edu/abs/2019ApJ...873..111I} {873, 111}

\bibitem[\protect\citeauthoryear{Jeong \& Kim}{Jeong \& Kim}{2021}]{2021kim}
Jeong M.,  Kim C.,  2021, in 2021 IEEE International Conference on Image
  Processing (ICIP). pp 739--743, \mn@doi{10.1109/ICIP42928.2021.9506364}

\bibitem[\protect\citeauthoryear{{Kalvankar}, {Pandit}  \&
  {Parwate}}{{Kalvankar} et~al.}{2020}]{2020arXiv200813611K}
{Kalvankar} S.,  {Pandit} H.,   {Parwate} P.,  2020, arXiv e-prints, \href
  {https://ui.adsabs.harvard.edu/abs/2020arXiv200813611K} {p. arXiv:2008.13611}

\bibitem[\protect\citeauthoryear{{Lotz}, {Primack}  \& {Madau}}{{Lotz}
  et~al.}{2004}]{2004AJ....128..163L}
{Lotz} J.~M.,  {Primack} J.,   {Madau} P.,  2004, \mn@doi [\aj]
  {10.1086/421849}, \href
  {https://ui.adsabs.harvard.edu/abs/2004AJ....128..163L} {128, 163}

\bibitem[\protect\citeauthoryear{Lupton, Gunn, Ivezic, Knapp, Kent  \&
  Yasuda}{Lupton et~al.}{2001}]{lupton2001sdss}
Lupton R.,  Gunn J.~E.,  Ivezic Z.,  Knapp G.~R.,  Kent S.,   Yasuda N.,  2001,
  \mn@doi [arXiv preprint astro-ph/0101420] {10.1117/12.457307}

\bibitem[\protect\citeauthoryear{Mittal, Soorya, Nagrath  \& Hemanth}{Mittal
  et~al.}{2020}]{mittal2020data}
Mittal A.,  Soorya A.,  Nagrath P.,   Hemanth D.~J.,  2020, \mn@doi [EARTH SCI
  INFORM] {10.1007/s12145-019-00434-8}, 13, 601

\bibitem[\protect\citeauthoryear{{Nishikawa-Toomey}, {Smith}  \&
  {Gal}}{{Nishikawa-Toomey} et~al.}{2020}]{nishikawa2020semi}
{Nishikawa-Toomey} M.,  {Smith} L.,   {Gal} Y.,  2020, arXiv e-prints, \href
  {https://ui.adsabs.harvard.edu/abs/2020arXiv201108714N} {p. arXiv:2011.08714}

\bibitem[\protect\citeauthoryear{{Ostrander}, {Nichol}, {Ratnatunga}  \&
  {Griffiths}}{{Ostrander} et~al.}{1998}]{1998AJ....116.2644O}
{Ostrander} E.~J.,  {Nichol} R.~C.,  {Ratnatunga} K.~U.,   {Griffiths} R.~E.,
  1998, \mn@doi [\aj] {10.1086/300627}, \href
  {https://ui.adsabs.harvard.edu/abs/1998AJ....116.2644O} {116, 2644}

\bibitem[\protect\citeauthoryear{{Sabour}, {Frosst}  \& {E Hinton}}{{Sabour}
  et~al.}{2017}]{sabour2017dynamic}
{Sabour} S.,  {Frosst} N.,   {E Hinton} G.,  2017, arXiv e-prints, \href
  {https://ui.adsabs.harvard.edu/abs/2017arXiv171009829S} {p. arXiv:1710.09829}

\bibitem[\protect\citeauthoryear{Scoville et~al.,}{Scoville
  et~al.}{2007}]{scoville2007cosmic}
Scoville N.,  et~al., 2007, \mn@doi [ApJs] {10.1086/516585}, 172, 1

\bibitem[\protect\citeauthoryear{Sersic}{Sersic}{1968}]{sersic1968atlas}
Sersic J.~L.,  1968, Cordoba, Argentina: Observatorio Astronomico, 1968

\bibitem[\protect\citeauthoryear{Sorrentino, Antonuccio-Delogu  \&
  Rifatto}{Sorrentino et~al.}{2006}]{sorrentino2006galaxy}
Sorrentino G.,  Antonuccio-Delogu V.,   Rifatto A.,  2006, \mn@doi [A&A]
  {10.1051/0004-6361:20065789}, 460, 673

\bibitem[\protect\citeauthoryear{Wang \& Xu}{Wang \&
  Xu}{2007}]{Wang2007Progressinastronomy}
Wang M.,  Xu X.,  2007, \mn@doi [Progress in astronomy]
  {10.3969/j.issn.1000-8349.2007.03.003}, 25, 215

\bibitem[\protect\citeauthoryear{Wang Lin-Qian \& Luo}{Wang Lin-Qian \&
  Luo}{2022}]{GMC2022Wang}
Wang Lin-Qian B.~Q.,  Luo A.-L.,  2022, \mn@doi [Astronomical Research &
  Technology] {10.14005/j.cnki.issn1672-7673.20210823.002}

\bibitem[\protect\citeauthoryear{{Willett} et~al.,}{{Willett}
  et~al.}{2013}]{2013MNRAS.435.2835W}
{Willett} K.~W.,  et~al., 2013, \mn@doi [\mnras] {10.1093/mnras/stt1458}, \href
  {https://ui.adsabs.harvard.edu/abs/2013MNRAS.435.2835W} {435, 2835}

\bibitem[\protect\citeauthoryear{{Xiang}, {Zhang}, {Tang}, {Zou}  \&
  {Xu}}{{Xiang} et~al.}{2018}]{2018xiang}
{Xiang} C.,  {Zhang} L.,  {Tang} Y.,  {Zou} W.,   {Xu} C.,  2018, \mn@doi [IEEE
  Signal Processing Letters] {10.1109/LSP.2018.2873892}, \href
  {https://ui.adsabs.harvard.edu/abs/2018ISPL...25.1850X} {25, 1850}

\bibitem[\protect\citeauthoryear{{Yao-Yu Lin}, {Liao}, {Huang}, {Kuo}  \&
  {Hsuan-Min Ou}}{{Yao-Yu Lin} et~al.}{2021}]{2021arXiv211001024Y}
{Yao-Yu Lin} J.,  {Liao} S.-M.,  {Huang} H.-J.,  {Kuo} W.-T.,   {Hsuan-Min Ou}
  O.,  2021, arXiv e-prints, \href
  {https://ui.adsabs.harvard.edu/abs/2021arXiv211001024Y} {p. arXiv:2110.01024}

\bibitem[\protect\citeauthoryear{{Zhang}, {Zou}, {Li}  \& {Chen}}{{Zhang}
  et~al.}{2022}]{2022RAA....22e5002Z}
{Zhang} Z.,  {Zou} Z.,  {Li} N.,   {Chen} Y.,  2022, \mn@doi [RAA]
  {10.1088/1674-4527/ac5732}, \href
  {https://ui.adsabs.harvard.edu/abs/2022RAA....22e5002Z} {22, 055002}

\bibitem[\protect\citeauthoryear{Zhao, Zhao, Chu, Jing  \& Deng}{Zhao
  et~al.}{2012}]{zhao2012lamost}
Zhao G.,  Zhao Y.-H.,  Chu Y.-Q.,  Jing Y.-P.,   Deng L.-C.,  2012, \mn@doi
  [RAA] {10.1088/1674-4527/12/7/002}, 12, 723

\bibitem[\protect\citeauthoryear{{Zhu}, {Dai}, {Bian}, {Chen}, {Chen}  \&
  {Hu}}{{Zhu} et~al.}{2019}]{zhu2019galaxy}
{Zhu} X.-P.,  {Dai} J.-M.,  {Bian} C.-J.,  {Chen} Y.,  {Chen} S.,   {Hu} C.,
  2019, \mn@doi [\apss] {10.1007/s10509-019-3540-1}, \href
  {https://ui.adsabs.harvard.edu/abs/2019Ap&SS.364...55Z} {364, 55}

\bibitem[\protect\citeauthoryear{van~der Maaten \& Hinton}{van~der Maaten \&
  Hinton}{2008}]{van2008visualizing}
van~der Maaten L.,  Hinton G.,  2008, Journal of Machine Learning Research, 9,
  2579

\makeatother
\end{thebibliography}




\appendix


\bsp	
\label{lastpage}
\end{document}